\documentclass{emulateapj}

\newcommand{\longobjname}{SDSS J094604.90+183541.8}
\newcommand{\objname}{J0946+1835}
\newcommand{\galfit}{GALFIT}
\newcommand{\lensmodel}{{\em lensmodel}}

\shorttitle{A Lensed Quasar at $z=4.8$}
\shortauthors{McGreer et al.}

\begin{document}

\title{SDSS J094604.90+183541.8: A Gravitationally Lensed Quasar at $z=4.8$}

\author{
Ian D. McGreer,\altaffilmark{1}
Patrick B. Hall,\altaffilmark{2}
Xiaohui Fan,\altaffilmark{1}
Fuyan Bian,\altaffilmark{1}
Naohisa Inada,\altaffilmark{3,4} 
Masamune Oguri,\altaffilmark{5} 
Michael A. Strauss,\altaffilmark{6}
Donald P. Schneider,\altaffilmark{7} and
Kara Farnsworth\altaffilmark{1}
}

\altaffiltext{1}{Steward Observatory, The University of Arizona, 
                 933 North Cherry Avenue, Tucson, AZ 85721--0065}
\altaffiltext{2}{Department of Physics and Astronomy, York University, 
                 4700 Keele St., Toronto, ON, M3J 1P3, Canada}
\altaffiltext{3}{Cosmic Radiation Laboratory, RIKEN, 2-1 Hirosawa, Wako,
                 Saitama 351-0198, Japan.} 
\altaffiltext{4}{Research Center for the Early Universe,
                 School of Science, University of Tokyo, Bunkyo-ku, Tokyo
                 113-0033, Japan.}
\altaffiltext{5}{National Astronomical Observatory of Japan, 2-21-1 Osawa,
                 Mitaka, Tokyo 181-8588, Japan.}
\altaffiltext{6}{Princeton University Observatory, Peyton Hall,
                 Princeton, NJ 08544, USA.}  
\altaffiltext{7}{Department of Astronomy and Astrophysics, The
                 Pennsylvania State University, 525 Davey Laboratory, 
                 University Park, PA 16802, USA.} 

\begin{abstract}
We report the discovery of a gravitationally lensed quasar identified 
serendipitously in the Sloan Digital Sky Survey (SDSS). The object, 
\longobjname, was initially targeted for spectroscopy as a luminous red 
galaxy, but the SDSS spectrum has the features of both a $z=0.388$ galaxy 
and a $z=4.8$ quasar. We have obtained additional imaging that resolves the 
system into two quasar images separated by 3.06\arcsec\ and a bright galaxy that
is strongly blended with one of the quasar images. We confirm spectroscopically 
that the two quasar images represent a single lensed source at $z=4.8$ with
a total magnification of 3.2, and we derive a model for the lensing galaxy. 
This is the highest redshift lensed quasar currently known. We examine the 
issues surrounding the selection of such an unusual object from existing data 
and briefly discuss implications for lensed quasar surveys.
\end{abstract}

\keywords{gravitational lensing -- quasars: individual (\longobjname)}

\section{Introduction}

Gravitationally lensed quasars offer the opportunity to study quasar host 
galaxies at a level of detail not otherwise accessible \citep{peng+06}, 
sightlines to study the physical extent of intervening absorption structures 
\citep{ellison+04}, and the means to constrain lens galaxy masses.
Lensed quasar statistics are highly useful probes of cosmology, and
time delays from individual lenses provide constraints on the Hubble 
constant \citep[for a review see][]{kochanek06}.
Several approaches have been adopted to construct large samples of lensed 
quasars.  One method is to obtain high-resolution images of a large number of 
sources to look for objects resolved into multiple images on small scales. 
This method has been applied at radio wavelengths by the CLASS survey, 
from which 22 lenses were found in a survey of $\sim16,000$ radio sources
\citep{myers+03,browne+03}. At optical wavelengths, the subarcsecond 
resolution required to identify most lenses has been achieved with HST 
snapshot surveys. \citet{maoz+93} found five lenses in a sample of 502 
quasars with $z>1$; more recently, extensive HST follow-up of SDSS quasars at 
$z>4$ yielded no lenses with $>0.1$\arcsec\ separation in a sample of 162 
objects \citep{richards+04,richards+06}. All of these searches can be 
considered more or less blind, in that little or no prior information was 
used to select lens candidates.

An alternative is to select likely candidates for lensed quasars directly 
from imaging data. This is the approach taken by the SDSS Quasar Lens Search 
(SQLS: \citealt{oguri+06,oguri+08}, \citealt{inada+08} and references 
therein, \citealt{kayo+09}), which has identified 40 new gravitational 
lenses through follow-up of quasars within the SDSS. Candidates are 
selected either by the presence of a nearby object with similar colors as 
the quasar, or by the quasar having image being resolved, a hint that it may 
represent the blend of multiple images at small separations.

Only a handful of $z>2.5$ quasars are currently known to be gravitationally 
lensed. Both blind and targeted surveys for lensed high redshift quasars are 
limited by the extreme rarity of the source population.  That no lenses were 
found in the survey of \citeauthor{richards+06} --- which included more than half
of the $z>4$ quasars known at the time --- is unsurprising given lensing rates 
of $<1\%$ found by surveys of quasars at lower redshifts 
\citep{inada+08,browne+03}; though the lensing probability increases with
redshift \citep{tog84} and magnification bias is expected to further enhance
the observed lensing rate at high redshift in flux-limited surveys 
(for discussions of this effect on SDSS quasars, see 
\citealt{wl02}~and~\citealt{comerford+02}). 
Targeted surveys such as the SQLS further depend 
on the initial selection method employed to identify high-redshift quasar 
candidates.  In the case of the SDSS, quasar targets at $z>2.5$ are required 
to have a point-like morphology, introducing a strong bias against 
small-separation lensed systems. 

We present the discovery of a $z=4.8$ lensed quasar drawn from 
the SDSS. This is the highest redshift lensed quasar currently known, and 
only the third at $z>4$, after BRI 0952-0115 at $z=4.426$ (1\arcsec\ 
separation, \citealt{mcmahon+92}) and PSS 2322+1944 at $z=4.12$ 
(1.49\arcsec\ separation, \citealt{carilli+02,riechers+08}). 
\longobjname\ (hereafter \objname) was initially targeted for spectroscopy 
as a luminous red galaxy and identified by strong quasar emission lines 
apparent in the SDSS spectrum. 

This paper is organized as follows: in Section~\ref{sec:obs} we show how the 
object was serendipitously identified from SDSS data and provide results from 
extensive photometric and spectroscopic follow-up that confirm the lensing 
hypothesis. We then provide results from modeling the lens galaxy 
(\S~\ref{sec:model}), briefly discuss the nature of the discovery of this 
object  (\S~\ref{sec:discuss}), and conclude with prospects for future 
searches for high-redshift quasar lenses (\S~\ref{sec:conclude}). Unless
otherwise noted, all magnitudes are on the AB system and corrected for 
Galactic reddening using the maps of \citet{schlegel}. We adopt a standard 
cosmology with parameters $H_0=70~{\rm km}~{\rm s}^{-1}~{\rm Mpc}^{-1},
\Omega_m=0.3, \Omega_\Lambda=0.7$.

\begin{figure}[!t]
 \epsscale{0.75}
 \plotone{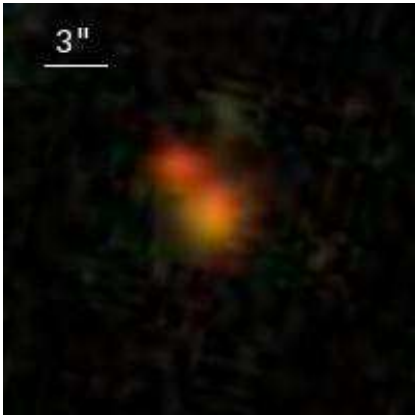}
 \caption{ 
  SDSS color image of \objname. The image is centered on the photometric
  object targeted as a galaxy; the strong blending from additional, redder
  objects is apparent.
 \label{fig:SDSSim}
 }
\end{figure}
\begin{deluxetable}{rcccc}
  \centering
  \tablecaption{SDSS Photometry}
  \tablewidth{0pt}
  \tablehead{ 
              \colhead{Object} & 
              \colhead{$g$} &
              \colhead{$r$} &
              \colhead{$i$} &
              \colhead{$z$} }
  \startdata
A & $>  23.4$          & $ 21.34 \pm  0.14$ & $ 19.52 \pm  0.12$ & $ 19.46 \pm  0.11$\\
B+G & $ 20.90 \pm  0.06$ & $ 18.87 \pm  0.01$ & $ 17.98 \pm  0.01$ & $ 17.66 \pm  0.02$\\
\multicolumn{2}{l}{\textit{GALFIT results:}} & & & \\
A$'$ & $>  23.4$        & $  21.42 \pm  0.04 $ & $  19.49 \pm  0.02 $ & $  19.53 \pm  0.05 $ \\
B$'$ & $>  23.4$        & $  20.46 \pm  0.05 $ & $  19.24 \pm  0.04 $ & $  19.30 \pm  0.17 $ \\
G$'$ & $  20.89 \pm  0.21 $ & $  18.95 \pm  0.06 $ & $  18.30 \pm  0.04 $ & $  17.80 \pm  0.08 $
  \enddata
  \label{tab:sdssphot}
  \tablecomments{The SDSS positions (J2000) are 09:46:04.90 +18:35:41.8 for A and 09:46:04.79 +18:35:39.7 for B+G. 
     For object A the values are from PSF magnitudes, while for object B+G
     the values are model magnitudes; the SDSS photometry is given in
     asinh magnitudes \citep{lupton+99}.
     Deblended photometry from \galfit\ is given in the last three rows
     and described in \S\ref{sec:obs_arc}; magnitudes are on the AB system.}
\end{deluxetable}
\begin{figure}[!t]
 \epsscale{1.2}
 \plotone{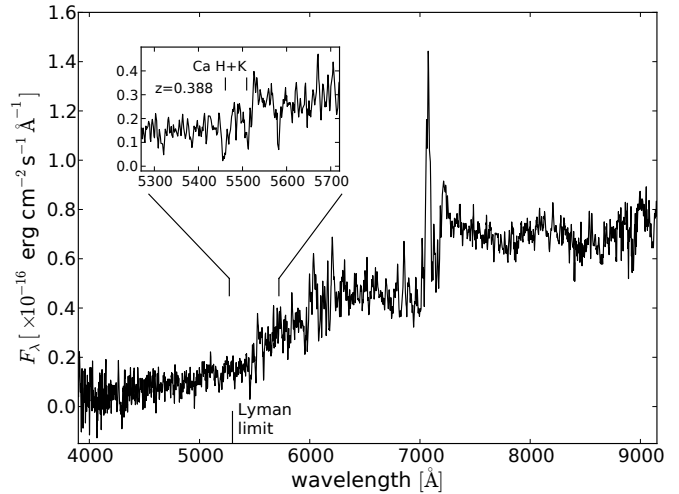}
 \caption{ 
  SDSS spectrum of \objname, smoothed with a 5 pixel boxcar function.
  The Ly$\alpha$ and \ion{N}{5} emission lines unambiguously show the
  signature of a $z=4.8$ quasar.  However, the level of flux blueward
  of Ly$\alpha$ (and in particular blueward of the Lyman limit) indicates
  excess light from another object; an expansion of the region around 5500\AA\ 
  shows that this object is a $z=0.388$ galaxy.
 \label{fig:SDSSspec}
 }
\end{figure}

\section{Observations and Data Reduction}\label{sec:obs}

\objname\ was targeted for spectroscopy by the SDSS as a 
Luminous Red Galaxy (LRG) according to the $gri$ color selection technique 
outlined in \citet{eisenstein+01}. As we will discuss in \S~\ref{sec:discuss}, this 
selection occurred because the foreground galaxy is significantly brighter than the 
background quasar images. 
In this section we present various observations of this system, beginning 
with the SDSS imaging and spectroscopy from which it was initially 
identified, followed by higher resolution and signal-to-noise ratio imaging, 
and finally spectroscopic observations  of the multiple components comprising 
the system. All of these data provide strong evidence in support of the lensing hypothesis.

\subsection{SDSS Imaging \& Spectroscopy}\label{sec:obs_sdss}

The SDSS is a photometric and spectroscopic survey covering nearly 
one-quarter of the sky \citep{york+00}. Images with the multi-CCD camera
\citep{gunn+98} are obtained in five optical bands ($ugriz$) and reduced
according to a photometric pipeline described in \citet{lupton+01}.
Details of the photometric calibration procedure are given in
\citet{fukugita+96} and \citet{smith+02}, and the astrometric calibration
is outlined in \citet{pier+03}. Photometric objects are selected for 
spectroscopy by various programs as detailed in \citet{stoughton+02}, 
including, among others, $\sim10^5$ quasar targets \citep{richards+02} and a 
similar number of Luminous Red Galaxy (LRG) targets \citep{eisenstein+01}.

\objname\ was imaged during the course of the SDSS main survey on
2005 Mar 10 with seeing of 1.2\arcsec, 1.2\arcsec, and 1.5\arcsec\ in
the $r$, $i$, and $z$ bands, respectively. The SDSS photometric pipeline 
identified two objects in this field: a stellar object, and a much brighter  
galaxy located $\sim2.4$\arcsec\ to the SW. As we will show, this galaxy 
is actually an unresolved blend of a low-redshift galaxy and a 
high-redshift quasar.  Although the galaxy is contaminated with quasar 
light, it was selected as an LRG spectroscopic target based on its $gri$ 
colors. Details of the SDSS photometry from Data Release 7 \citep[DR7,][]{dr7} 
for the two objects are given in Table~\ref{tab:sdssphot}.

SDSS spectra are acquired with a multi-fiber spectrograph, with each fiber
subtending a 3\arcsec\ diameter region on the sky. The spectra span
3800-9200\AA\ at a resolution of $\sim1800$; the wavelength and flux
calibration of the spectra are described in \citet{stoughton+02}.
A spectrum was obtained on 2006 Jan 24 with 1.3\arcsec\ seeing.  The 
SDSS fiber was centered on the position derived for the B+G object.
The spectrum clearly shows the emission line features of a $z=4.8$ quasar 
(Figure~\ref{fig:SDSSspec}). The spectrum also shows the Ca H+K absorption 
features of a $z=0.388$ galaxy. The automated spectroscopic pipeline 
\texttt{spectro1d} \citep{stoughton+02} classified the object as a quasar 
with low confidence (56\%) and assigned a redshift $z=4.806$. 

The complex nature of the spectrum became apparent from manual inspection
during construction of the SDSS DR7 quasar catalog \citep{dr7qso}.
One author (DPS) identified 1083 spectra from DR6 and DR7 with quasar
classifications but with questionable redshifts.  From among those objects
another author (PBH) identified six quasar/galaxy spectral blends
whose SDSS images were consistent with a lensing hypothesis.
Of those six, one had already been been studied in the SQLS and found 
to be a quasar-galaxy pair, one was determined to be a quasar next to a 
$z=0.0792$ galaxy, three are being followed up further, and the sixth is the 
object reported here.

\begin{figure}[!t]
 \epsscale{1.2}
\plotone{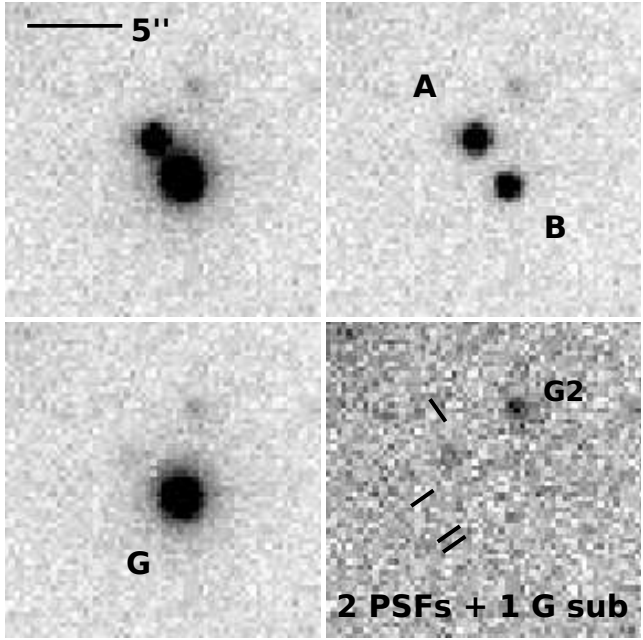}
 \caption{ 
  ARC SPIcam imaging of \objname\ with 0{\farcs}9 seeing. 
  North is up and east is left.
  The upper left panel shows a cutout image from the $i$ band. 
  Three components have been fit with \galfit; the upper right panel shows
  the residual image after the model galaxy has been removed,
  while the lower left panel shows the residual image after the two model
  PSFs have been removed.
  The lower right  panel 
  shows the residual image with all model components subtracted, with lines
  whose intersections indicate the positions of each of the components.
  A faint galaxy ($i=22.1$) remains visible to the NW, as well as a faint
  excess to the east.
 \label{fig:APOims}
 }
\end{figure}
\begin{figure}[!t]
 \epsscale{1.2}
 \plotone{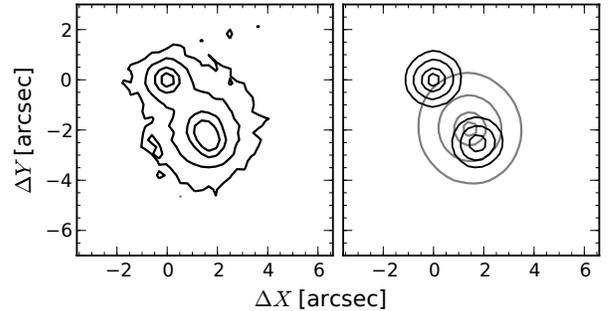}
 \caption{ 
  Contour plots of the ARC $i$-band imaging.
  Left panel: Total flux contours.
  Right panel: Contours of flux for the fitted components,
  including the two PSFs (black lines) and the model galaxy (gray lines).
  All contours are drawn at (19.6, 20.2, 21.4, 22.6)~mag~arcsec$^{-2}$,
  and positions are relative to A.
 \label{fig:APOcontours}
 }
\end{figure}
\begin{deluxetable*}{crrrrrrr}
  \centering
  %APJREMOVE
  \tabletypesize{\footnotesize}
  \tablecaption{Relative Astrometry and Photometry}
  \tablewidth{0pt}
  \tablehead{ 
              \multicolumn{3}{c}{} &
              \multicolumn{3}{c}{UH88} &
              \multicolumn{2}{c}{ARC} \\
              \colhead{Component} & 
              \colhead{$\Delta $X (\arcsec)} &
              \colhead{$\Delta $Y (\arcsec)} &
              \colhead{$V$} &
              \colhead{$R$} &
              \colhead{$I$} &
              \colhead{$i$} &
              \colhead{$z$} }
  \startdata
A & $  0.000 \pm  0.003$ & $  0.000 \pm  0.003$ &  & $21.05 \pm 0.02$ & $19.90 \pm 0.01$ & $19.76 \pm 0.01$ & $19.71 \pm 0.02$ \\
B & $  1.756 \pm  0.005$ & $ -2.503 \pm  0.005$ &  & $21.74 \pm 0.07$ & $20.00 \pm 0.04$ & $20.12 \pm 0.03$ & $19.82 \pm 0.05$ \\
G & $  1.453 \pm  0.005$ & $ -1.942 \pm  0.008$ & $19.37 \pm 0.06$ & $18.75 \pm 0.02$ & $18.06 \pm 0.02$ & $18.34 \pm 0.01$ & $17.96 \pm 0.02$ 
  \enddata
  \label{tab:imagefit}
  \tablecomments{The reference position for component A is 09:46:04.90 +18:35:41.85 (J2000), with 
    a systematic error in mapping to SDSS astrometry of 0\farcs06 in each 
    coordinate. All positions are derived from the ARC $i$-band imaging; 
    positive X and Y represent west and north, respectively. Magnitudes are
    based on PSF models for A and B and S\'ersic model for G, as calculated
    by \galfit.}
\end{deluxetable*}
\begin{deluxetable*}{rrrrrr}
  \centering
  \tablecaption{Parameters of the galaxy profile}
  \tablewidth{0pt}
  \tablehead{ \colhead{} & 
              \colhead{$V$} &
              \colhead{$R$} &
              \colhead{$I$} &
              \colhead{$i$} &
              \colhead{$z$} }
  \startdata
    $R_e ({\rm \arcsec})$ & $  2.59 \pm  0.26 $  & $  1.56 \pm  0.07 $  & $  1.89 \pm  0.11 $  & $  1.24 \pm  0.03 $  & $  1.33 \pm  0.05 $ \\
                      $e$ & $  0.29 \pm  0.05 $  & $  0.16 \pm  0.02 $  & $  0.13 \pm  0.02 $  & $  0.13 \pm  0.01 $  & $  0.13 \pm  0.02 $ \\
    $\theta_e ({}^\circ)$ & $  28.8 \pm  6.5 $  & $  14.2 \pm  5.4 $  & $  26.7 \pm  7.0 $  & $  16.6 \pm  5.0 $  & $  24.9 \pm  8.6 $ 
  \enddata
  \label{tab:galaxyfit}
  \tablecomments{All measurements taken from \galfit\ results based on fitting 
      a S\'ersic profile with $n=4$. The parameters are the effective radius 
      ($R_e$), ellipticity ($e \equiv 1-b/a$), and position angle of the 
      ellipticity ($\theta_e$, measured east of north).}
\end{deluxetable*}
\begin{deluxetable}{lrrr}
  \centering
  %APJREMOVE
  \tabletypesize{\footnotesize}
  \tablecaption{Multi-epoch $i$-band Photometry}
  \tablewidth{0pt}
  \tablehead{ 
              \colhead{Source} & 
              \colhead{Epoch} & 
              \colhead{A} &
              \colhead{B} }
  \startdata
 SDSS & 2005.19 & 19.49 & 19.24 \\
  ARC & 2009.85 & 19.76 & 20.12 \\
 UH88\tablenotemark{a} & 2010.04 & 19.82 & 19.92 \\
  \enddata
  \label{tab:epochobs}
  \tablecomments{Photometry for the two quasar images over three observational
epochs.}% Systematic uncertainties are $\sim0.2$ mag for the SDSS and
%$\sim0.1$ mag for ARC/UH88.}
  \tablenotetext{a}{UH88 $i$-band magnitudes are derived from the $I$-band 
imaging after computing a color difference from the observed SED of
$I-i=0.08$.}
\end{deluxetable}

\subsection{ARC Imaging}\label{sec:obs_arc}

\objname\ was observed with the SPIcam imager on the ARC 3.5m telescope at 
Apache Point Observatory on 2009 Nov 6.  The $2048^2$ backside-illuminated 
SITe CCD was binned by a factor of two along each axis during readout, 
resulting in a pixel scale of 0.28\arcsec\ over the 4.8\arcmin\ field of 
view. The seeing was 0.9\arcsec\ and conditions were photometric. 
Two dithered exposures of 240s each were obtained in the $i$-band and two 
dithered exposures of 300s each were obtained in the $z$-band. The images 
were reduced using standard IRAF\footnote{IRAF is distributed by the National 
Optical Astronomy Observatories, which are operated by the Association of 
Universities for Research in Astronomy, Inc., under cooperative agreement with 
the National Science Foundation.} 
routines called from Pyraf\footnote{Pyraf is a product of the Space Telescope 
Science Institute, which is operated by AURA for NASA.} 
scripts. After applying a bias correction, supersky images for both filters 
were constructed by combining science images from other fields obtained just 
before the observations of \objname; a total of seven $i$-band and 23 
$z$-band images were used to construct the flats. Both astrometric 
registration and photometric calibration of the images were accomplished 
using bright SDSS stars within the field.

We derived positions and fluxes for the various components using
the \galfit\ code \citep{galfit}.  For both the $i$ and $z$ images,
approximate positions and fluxes obtained from examining the images
with IRAF were used as initial conditions for the fitting. An empirical 
PSF was constructed from a set of 10 bright stars within the field.
The fitting was then performed using two PSFs to represent the quasar
images and a S\'ersic model for the galaxy. Results of the
fitting are shown in Figure~\ref{fig:APOims} and Table~\ref{tab:imagefit}. 
For the $i$-band image, the separation between the quasar images is 
3.06\arcsec\ and the flux ratio is $f_{B}/f_{A} = 0.7$.
Isophotal contours for the model components derived from the $i$-band imaging
are displayed in Figure~\ref{fig:APOcontours}, showing the strong blending
between B and G. Table~\ref{tab:galaxyfit} shows parameters for the S\'ersic
profile derived by \galfit\ for the lens galaxy.

A faint galaxy is also seen $\sim5$\arcsec\ to the northwest 
(labeled G2 in Figure~\ref{fig:APOims}).
This galaxy is not detected by the SDSS pipeline, but we were able to extract
aperture photometry in the $g$, $r$, and $i$ bands, and obtain a flux of
$r=22.0 \pm 0.1$ and colors $g-r=1.3 \pm 0.3$ and $r-i=0.3 \pm 0.2$. Photometry 
from the ARC imaging yields $i=22.1 \pm 0.1$ and $i-z=0.3 \pm 0.2$. These colors 
are inconsistent with an object at the redshift of the quasar, indicating that it is not 
an additional lensed image. However, this faint galaxy may be associated with 
the bright lensing galaxy.
There also appears to be a flux excess (considerably fainter than
galaxy G2) to the east of quasar image A in the SPIcam $i$ band data.

\subsection{UH88 Imaging}\label{sec:obs_uh88}

Additional imaging data of \objname\ were obtained on 2010 Jan 16 using the
University of Hawaii 2.2m telescope (UH88) with the Tektronix $2048^2$ CCD 
camera, with a pixel scale of 0.22\arcsec. Conditions were clear and 
photometric with 0.8-1.0\arcsec\ seeing. Images were obtained in the $VRI$ 
bands; each band received four exposures of 120s each. Images were reduced 
using standard IRAF routines and photometric calibration was provided by the 
SA92 field. Magnitudes were derived on the Vega system and converted to
AB using $V_{\rm AB}=V_{\rm Vega} + 0.00, R_{\rm AB}=R_{\rm Vega} + 0.20,~
{\rm and}~I_{\rm AB}=I_{\rm Vega} + 0.45$.

The UH88 images were also fit by three components using \galfit\ using a
similar process to that of the ARC imaging. Results of fitting the UH88 
images are given in Tables~\ref{tab:imagefit}~and~\ref{tab:galaxyfit}.
Overall, there is a high level of consistency between the positions derived 
by \galfit\ in all of the imaging bands, and also between the UH88 and
ARC imaging. 

The galaxy G2 is detected in $V$, $R$, and $I$. As with the SPIcam $i$-band
image, excess flux appears to the east of quasar image A in the UH88
$I$ band image, indicating a real detection. There is a $\sim1.4\sigma$ excess 
at this location in the $V$ band; this object is also likely to be a low-redshift galaxy.

\subsection{Imaging Summary}\label{sec:obs_imgsummary}

We have presented results from three epochs of imaging for this 
system, summarized in Table~\ref{tab:epochobs}.
There are some apparent differences when comparing the
observations.
We first note that the quoted photometric errors 
(Tables~\ref{tab:sdssphot}~and~\ref{tab:imagefit}) 
are purely statistical errors
from \galfit\ and do not reflect the uncertainties arising from the fitting itself and
from the photometric calibration.
The fitting uncertainties are particularly important for the SDSS imaging, which 
has the poorest
resolution and the largest pixel size, hampering robust image decomposition.
We have derived deblended photometry from the SDSS images by using \galfit\ with 
some constraints to improve the fitting, namely, we fix the positions of the three 
components to the ARC-derived positions, and fix the galaxy fluxes to be within 
0.1 mag of the ARC values\footnote{For the $g$ and $r$ fluxes we used the $V$ and 
$R$ values from \S\ref{sec:obs_uh88}, and relaxed the constraint to $\pm 0.25$ mag.}.
The results of this fitting are shown in Table~\ref{tab:sdssphot}.
We also experimented with unconstrained fitting of the SDSS images and had poor
results. In particular, offsetting the initial positions by just a few pixels
generally results in a fit with only two components, component B effectively
being lost. This highlights the difficulty of fully characterizing a system
such as this from SDSS imaging alone.

The colors of component B are slightly redder than those of component A in
both the ARC and UH88 imaging, but not in the SDSS imaging. The ARC colors are 
$(i-z)_A = 0.05 \pm 0.02$ and 
$(i-z)_B = 0.30 \pm 0.06$;
on the other hand, the SDSS
colors are $i-z \approx 0$ for both components.
The redder color of component B may be due to greater extinction from the lens
galaxy; however, we argue against differential reddening of the two images
in \S~\ref{sec:obs_mmt}.
Some of the observed color 
differences between the two components are due to emission line differences
of the quasar spectra, discussed in \S~\ref{sec:obs_mmt}. The remainder
are likely due to the difficulty of decomposing the image properly, and are
indicative of systematic errors in the photometry greater than the
quoted statistical errors.

Component B is significantly brighter in the SDSS imaging
than in either the ARC or UH88 imaging. While the fluxes measured for component
A are roughly constant, component B appears to have faded by $\sim0.8$ mag
between the SDSS and ARC/UH88 epochs. While we have already concluded that
systematic uncertainties in the photometry may be large ($\ga 0.1$ mag),
from experimentation with \galfit\ we consistently find that B is brighter than 
A in the SDSS imaging. % visible by eye?
The relative fluxes of the two components may have changed due to
intrinsic variability of the source quasar observed with a time delay between
the two images, and/or differential microlensing of the two images. Component
B is close to the center of the lens galaxy and its observed flux decrease is
quite large; therefore, microlensing would seem to be the most likely 
explanation.

\subsection{MMT Spectroscopy}\label{sec:obs_mmt}

Spectroscopic observations were obtained on 2009 Nov 11 with the
Red Channel instrument on the MMT 6.5m.  These observations were
nearly contemporaneous with the ARC imaging, occuring only five
days later. A single 1800s exposure was obtained with a 1.5\arcsec\ 
slit and the 270 mm$^{-1}$ grating centered at 7500\AA.  This provides
a dispersion of 3.52 \AA~pix$^{-1}$ at a spectral resolution of $R\sim600$.
The FWHM of unresolved emission lines from wavelength calibration exposures 
is 3.6 pix. The spatial pixel scale is 0.297~arcsec~pix$^{-1}$.
The seeing was 1\arcsec\ and thin clouds were passing overhead during
the observation. The instrument was rotated to PA=36.6$^\circ$ in order
to align the two components from the SDSS imaging in the slit.

Lamp exposures were obtained immediately subsequent to the science
observation; however, a standard star observation was not obtained
until the following night, using an identical spectrograph configuration
but with somewhat better seeing. All images were reduced with standard
IRAF routines, including a bias correction, flat fielding from a
continuum lamp, wavelength calibration from HeNeAr lamps, and relative 
flux calibration from the standard star BD+284211 obtained on 2009 Nov 12.
The science image was corrected for cosmic rays using the L.A. Cosmic
routines \citep{lacosmic}.

\begin{figure}[!t]
 \epsscale{1.1}
 \plotone{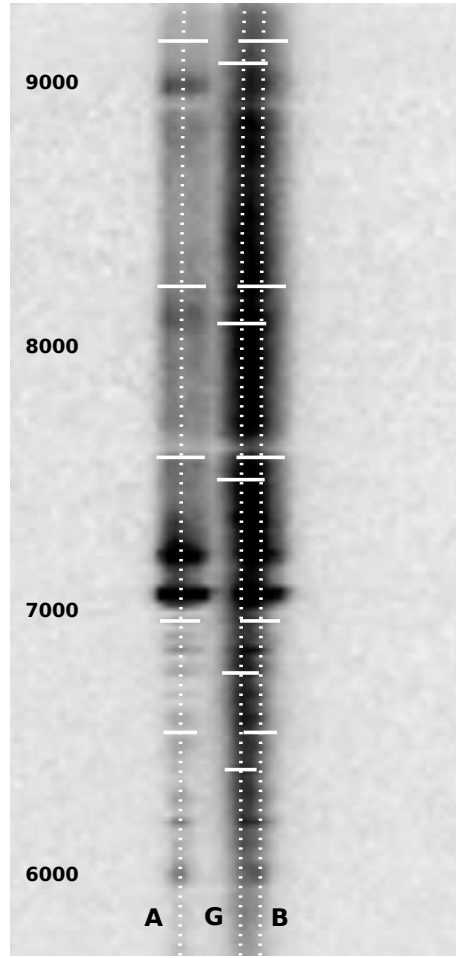}
 \caption{Two-dimensional spectra from MMT Red Channel observations. 
   The spectral image has been binned by a factor of 8 along rows for 
   display purposes (the spectral resolution is $R\sim600$). Approximate 
   wavelength positions along the dispersion axis (rows) are labeled. The 
   traces for each component derived from Gaussian fitting are shown as 
   dashed lines, while the FWHM of the fits are shown at various intervals 
   with a horizontal line. The A spectrum is relatively unblended with the other 
   two, while B and G are strongly blended.
 \label{fig:MMTtrace}
 }
\end{figure}
\begin{figure*}[!t]
 \epsscale{1.1}
 \plotone{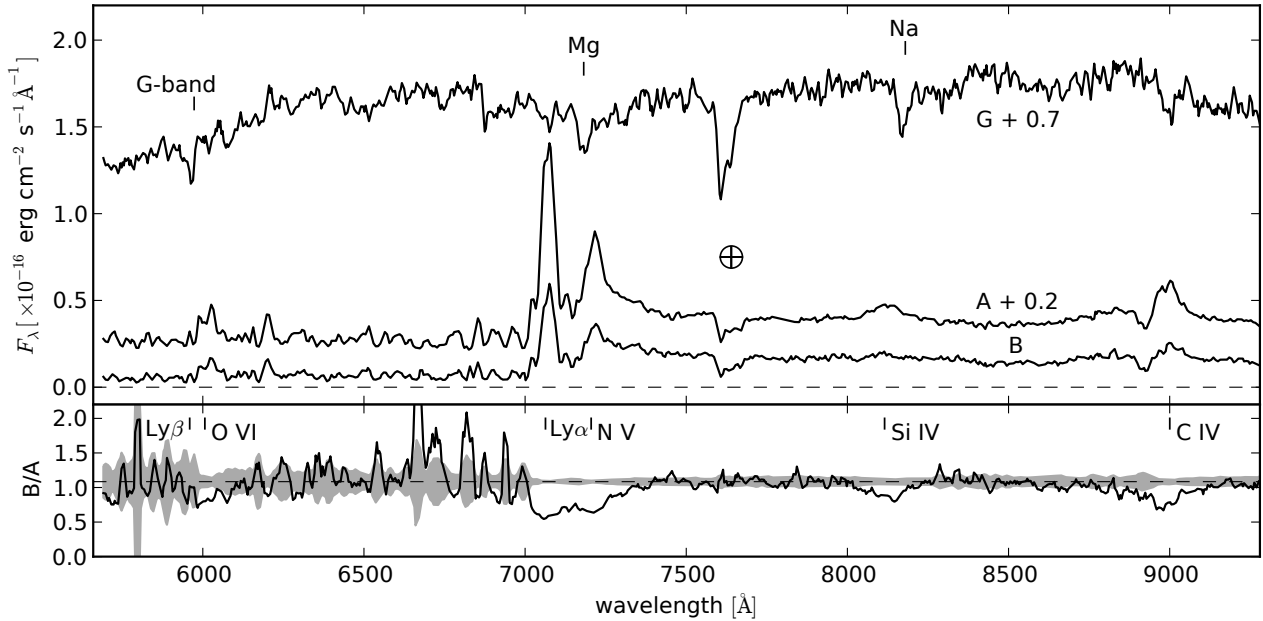}
 \caption{ 
  MMT Red Channel spectra of the components of \objname. 
  The upper panel shows the spectra extracted according to the procedure 
  described in \S\ref{sec:obs_mmt}. The spectra have been smoothed to
  a dispersion of 10.6\AA, and scaled to the $i-$band fluxes derived from 
  the image fitting of each component (Table~\ref{tab:imagefit}).
  Constant offsets have been added to the spectra of A and G for
  display purposes. The striking similarity of the  spectra for A and B, 
  including the line profiles and absorption blueward of \ion{N}{5} 
  and \ion{C}{4}, provides strong evidence for the lensing 
  hypothesis. Prominent elliptical galaxy absorption features are marked 
  above the G spectrum, using the redshift for the galaxy derived from the 
  Ca absorption seen in the SDSS spectrum (Figure~\ref{fig:SDSSspec}).
  The A and B spectra have $S/N \sim 30~{\rm pix}^{-1}$ on the red side of 
  Ly$\alpha$ and $S/N \sim 10~{\rm pix}^{-1}$ on the blue side.
  The lower panel shows the ratio of the B and A spectra after smoothing, 
  with no scaling of  the extracted spectra applied.  The locations of the
  prominent quasar emission lines are indicated. 
  The median ratio is $\sim1.1$ in the continuum regions, but significantly
  lower around the emission lines. 
  Uncertainties of $\pm 1\sigma$ in the ratio are shaded around the 
  median value.
  Note that this is the observed flux ratio,
  and may differ from the actual ratio of the two components due to 
  relative slit losses.
 \label{fig:MMTspec}
 }
\end{figure*}

The 2D spectra are strongly blended; in particular, components B and G are 
separated by only $\sim2$ pixels (0{\farcs}6) perpendicular to the dispersion 
direction (rows in the MMT image, see Figure~\ref{fig:MMTtrace}).  We adopted 
a heuristic approach to extraction of the spectra with all three components 
deblended, similar to that of \citet{pindor+04}.  First, the three spectra 
were treated as Gaussians fit independently to each row of the image, using 
a modified Levenberg-Marquardt algorithm from the SciPy library.  We found 
that allowing all Gaussian components to vary and solving for all nine 
parameters led to clearly unacceptable results, even with reasonable initial 
conditions. We thus found it necessary to apply strong constraints to the 
Gaussian parameters in order to produce acceptable fits. We assumed that the
standard star observation provides a model of the spectral trace on a 2D 
image, and then used this model to extract the blended science spectra, according to the following process:

\begin{enumerate}

\item A template trace for the 2D spectra was obtained from the high S/N 
      observation of the standard star. This trace was used to derive the 
      variation of the position and width of the spectrum along the image 
      rows. The trace positions were fit with a 7th order polynomial, but 
      they vary by only $\sim1$ pixel across the full image.

\item Along each row of the science image, a Gaussian with three free 
      parameters was fit to the leftmost columns of component A (those least 
      affected by blending from G).  This fit was compared to the template 
      trace, and used to derive a translation of the template along the image 
      columns to match the trace of A, as well as to derive a scale factor 
      for the widths corresponding to the difference in seeing between 
      science and standard star observations. This scale factor combined with 
      the template widths provides an estimate of the widths of the point 
      source components A and B.

\item The fit to A from step \#2 was subtracted from the image. The same 
      fit was then translated along columns to the position of B (as derived 
      from the imaging) and subtracted again. The residual image is an 
      approximation to the spectrum of G alone.  Using the template trace 
      again, fitting was performed on the residual image to derive the width 
      of the galaxy spectrum along all rows.

\item The previous steps provide estimates for the positions and widths of 
      each spectral component based on expected positions derived from the 
      imaging observations.  However, the slit was likely not aligned perfectly 
      on the center of each component.  We then used the derived fits as 
      initial conditions to a fit on the full 2D image.  For this fit, the 
      centers of the traces for each of the three components were offset by 
      random amounts of up to $\pm1$ pixel along the columns.  Fitting across 
      the full image was then repeated 500 times with different initial 
      offsets, allowing only the trace positions to vary. In other words, 
      copies of the template trace derived from the standard star spectrum 
      were varied in position until the best alignment with the three 
      components was found. The positions from the best $\chi^2$ among the 
      repeated fits were then adopted as the best-fit positions for the 2D 
      spectra.

\item Finally, with all positions and widths fixed, a fit was performed 
      allowing only the amplitudes to vary along each row. The integrated 
      Gaussians for the resulting amplitudes and widths were taken to 
      represent the 1D spectra for each of the components.

\end{enumerate}

Figure~\ref{fig:MMTspec} shows the extracted spectra for the three 
components. The spectra have been scaled such that the synthetic $i$-band 
fluxes derived from the spectra match the $i$-band fluxes from the ARC 
imaging.

Several features of the spectra are worth noting. First, both the emission 
and absorption features of A and B show remarkable agreement, providing clear
evidence that they represent lensed images of a single quasar at a redshift
$z=4.8$. This cannot be due to incorrect extraction of the 2D spectra, as
the Gaussians used to extract these two components are cleanly separated. 
While the line profiles of A and B are almost indistinguishable, the ratio
of the two spectra shows that the emission line to continuum ratios are 
different. Component B is well within the effective radius of the galaxy,
and differential microlensing of the continuum and broad line regions of
B could explain this effect \citep[e.g.,][]{richards+04a}. Alternatively, 
spectral variability could produce the discrepant line to continuum ratios, 
and cannot be ruled out by our single-epoch spectra. While the relative 
fluxes of the extracted spectra for the two components are uncertain due to 
possible differential slit losses, and the extraction was hampered by the 
strong blending of the B and G spectra, we are confident that the observed 
difference between the line to continuum ratios is not an artifact of our 
data. We have experimented with a large number of fitting parameters for 
extracting the 2D spectra, and in no viable fit does the B spectrum appear 
to be simply scaled from the A spectrum; in other words, the line ratios of 
the two components are always found to be significantly different from the 
continuum ratios.

In Section~\ref{sec:obs_imgsummary} we noted color differences between the
two components of $\Delta(i-z) \approx 0.25$ mag, and considered whether
differential reddening was affecting the observed colors.
If one of the images, namely component B, was strongly affected 
by extinction from the lens galaxy, this would appear in the spectra as a 
difference in the shapes of the continua measured for the two components. 
Instead, the ratio of the continua measured for the two components is flat across
$\sim 2000$\AA\ of the observed wavelength range. However, some color differences 
should be introduced by the observed emission line differences.
We derive synthetic $i$- and $z$-band magnitudes from the MMT spectra and find 
that the emission line differences contribute $\sim0.13$ mag to the redder color 
of component B. This reduces the net color difference between the two components 
from the ARC imaging to $\sim0.12$ mag, which is within the uncertainties.

The absorption features in the galaxy spectrum are exactly as expected for 
a luminous elliptical galaxy at $z=0.388$, as derived from the Ca absorption 
features in the SDSS spectrum.  This provides further confidence that the 
extraction was successful, as neither do the galaxy absorption features 
appear in the quasar spectra, nor do the quasar emission features appear in 
the galaxy spectrum.

We determine the quasar redshift to be $z=4.8$ from the peak of the 
relatively narrow Ly$\alpha$ line, using the spectrum obtained for component A. 
The \ion{C}{4} line has significant absorption of its blue wing. A single Gaussian fit to the 
\ion{C}{4} line with the absorbed region masked results in a lower redshift of 
$z=4.77$; however, the broad component of the \ion{C}{4} line is often 
blueshifted with respect to the systemic redshift \citep{richards+02b}. We 
fit a power law continuum to the A spectrum and three Gaussians to the 
\ion{C}{4} line, one for a broad component, one for a narrow component, and 
one for the strong absorption. From this fit, the redshift of the narrow 
component is $z=4.80$, in good agreement with the Ly$\alpha$ redshift, while 
the broad component is blueshifted by $\sim2000~{\rm km}~{\rm s}^{-1}$.

While the quasar spectra do show absorption features, particularly between 
Ly$\alpha$ and \ion{N}{5} and blueward of \ion{C}{4}, the width of 
the \ion{C}{4} absorption is not strong enough to meet the criteria generally 
adopted for Broad Absorption Line (BAL) quasars \citep{weymann+91},
though it has a non-zero Absorption Index (AI = 140~km~s$^{-1}$) as defined
by \citet{hall+02} and may be a mini-BAL. This type of absorption appears in
a minority of quasars and its identical shape in the spectra of both quasar
images is strong evidence that they are lensed images of a single source as 
opposed to a binary quasar.

\section{Lens Modeling}\label{sec:model}

\begin{table}
  \caption{Lens Modeling Results}\label{tab:lensmodel}
  \begin{center}
   \vspace{-10pt}
   \begin{tabular}{rl}
    \hline
     $R_E({\rm\arcsec})$ & $1.34 \pm 0.03$\\
     $e$ & $0.41 \pm 0.04$ \\
     $\theta_e({^\circ})$ & $-49.0 \pm 1.4$\\
     $\Delta t({\rm days})$ & $142.7~\pm~6.8$\\
     $\mu_{\rm tot}$ & 3.2 \\
     \hline
   \end{tabular}
  \end{center}
\end{table}

We model the system as a Singular Isothermal Ellipsoid (SIE) lens generating 
two images of the source quasar. We use the  \lensmodel\ code
\citep{lensmodel} to calculate the properties of both the lensing galaxy and 
the quasar. The positions and fluxes of each of the quasar images and the 
position of the lens galaxy obtained from the \galfit\ fitting of the ARC 
$i$-band images are used to provide a total of 8 constraints. We solve for 
the position and flux of the source quasar, as well as the Einstein radius 
($R_{\rm E}$), ellipticity ($e$, defined as in Table~\ref{tab:galaxyfit}), 
position angle ($\theta_e$), and position of the lensing galaxy.  As the model 
has zero degrees of freedom, the 
\lensmodel\ solution converges to $\chi^2\sim0$.  The results derived from 
the ARC $i$-band imaging are given in Table~\ref{tab:lensmodel}, along with 
$1\sigma$ uncertainties ($\Delta\chi^2=1$). 
When fitting the lens model, positional uncertainties were increased by 50\% 
and the flux uncertainties for all components were set to 5\% in order 
to account for systematic uncertainties in deblending.
The known 
redshift of the lensing galaxy ($z=0.388$) enables us to estimate a time 
delay of 143 days.

There are some discrepancies between the galaxy light profile as measured
from the imaging (Table~\ref{tab:imagefit}) and the derived mass model
(Table~\ref{tab:lensmodel}). If these discrepancies are real, one possibility
is that  they are due to shear \citep[e.g.,][]{keeton+98} from the nearby 
galaxies that were not included in the model.

\section{Discussion}\label{sec:discuss}

Quasars in the SDSS have received extensive follow-up to identify 
lensed objects. The SQLS identifies candidate lensed quasars from the 
spectroscopically confirmed quasar sample using selection criteria outlined in 
\citet{oguri+06}. First, known quasars that are resolved in the imaging may 
represent two or more lensed images at small separations that are blended 
into single photometric objects within the SDSS database (morphological 
selection). Second, large separation lenses are selected by searching for 
nearby objects with similar colors as the quasar (color selection). 
Using these methods,
\citet{inada+09} recently identified five $z>2.2$ lensed quasars in the 
SDSS. Two are $z>3$ quasars targeted by the high redshift quasar selection 
criteria used by the SDSS \citep{richards+02}; for both objects, the two quasar 
images are separated by $>1.5$\arcsec\ and were successfully deblended by the
photometric pipeline.  \objname\ bears a strong resemblance to the $z=3.6$ 
lensed quasar found by \citet{johnston+03}, which was also a galaxy targeted as an 
LRG by the SDSS and shown to be a high redshift lensed quasar by the 
presence of strong, broad emission lines in the SDSS spectrum. 

It is instructive to consider whether \objname\ could have been selected as a 
lensed quasar candidate in the SQLS even if the lensing galaxy had not been 
targeted for spectroscopy by the SDSS.  The SDSS registers \objname\ as two 
photometric objects, one a point-like object representing component A, and the 
other a resolved object representing a blend of B and G.  The object representing 
component A is classified as stellar and its colors are relatively unaffected 
by the lensing galaxy, and thus it could have been considered as a quasar 
target by the SDSS.  In fact, the object is well within the color selection 
criteria used by the SDSS to target $z>4.5$ quasars \citep{richards+02}, with 
a dereddened PSF magnitude of $i=19.52$ and colors $r-i=1.82$ and $i-z=0.06$. 
It was not considered as a quasar target because it has the photometric flag 
DEBLEND\_NOPEAK set in multiple bands, indicating that no peak was found in 
the deblending. This means that the deblending, and thus the colors, are 
questionable, and the SDSS quasar target selection algorithm excludes such 
objects from its color selection.

Had component A not been flagged DEBLEND\_NOPEAK, it would have been a 
primary quasar target in the SDSS.  Once identified as a quasar, it would 
be further considered as a lens candidate by the SQLS. Component A has a 
stellar classification in the SDSS and its PSF likelihood is high enough 
(\texttt{star\_L}(r)=0.06 and \texttt{star\_L}(i)=0.07) to fail to meet
the morphological selection criteria of the SQLS \citep{oguri+06}.  In 
addition, the strong blending of the quasar and galaxy flux in the B+G 
object causes significant color differences between the two SDSS objects 
($(r-i)_{A} - (r-i)_{B+G} = 0.93$), such that it would also fail to satisfy the 
color selection criteria. Thus, in the hypothetical (though generally 
expected) case that the stellar SDSS object was targeted as a quasar instead 
of the galaxy object being targeted as an LRG, this object would not have 
been identified as a lens candidate. It was fortuitous that the lensing 
galaxy was the object targeted for spectroscopy, as the blended quasar and 
galaxy features in the spectrum made the nature of the object evident.  
Furthermore, since the B+G object was spectroscopically identified as a 
quasar but clearly has the morphology of a galaxy, \objname\ will be included
in the SQLS DR7 additional sample by modified morphological selection
\cite[see Section 5 of][]{inada+08}.

Many of the difficulties encountered with selecting this object as a 
gravitational lens candidate arise from the bright lensing galaxy. The galaxy 
effectively hides the second quasar image, both in terms of detecting a 
second PSF image, as well as the blending of the galaxy and quasar colors. 
The galaxy also affected the well-separated quasar image by inducing additional 
photometric errors.  High-redshift quasars have extremely red colors and are 
detected in only a few optical bands, and it is easy to imagine that similar 
lensed quasars may be hiding behind bright  lensing galaxies. One way to find 
such objects is in the spectra of galaxy targets, as was the case here. This 
method has been used explicitly to find lensed background galaxies in the 
spectra of LRGs by the Sloan Lens ACS Survey \citep[SLACS;][]{bolton+06}.

\section{Conclusions}\label{sec:conclude}

We have identified the highest known redshift lensed quasar. 
\objname\ consists of a lens galaxy at $z=0.388$ and a pair of quasar images 
produced from a source at $z=4.8$. An object representing a blend of the 
lensing galaxy and one of the quasar images was targeted as a luminous red 
galaxy by the SDSS and classified as a quasar based on the spectrum. 
Subsequent observations show that the quasar images are separated by 
3\arcsec\ and have a relatively low total magnification of 3.2.  
Comparison of photometry from three epochs indicates that there is variability of the 
two quasar images. Deblended spectra of the three components show that the 
two images represent a single quasar, and suggest that microlensing may affect 
the relative line equivalent widths, though spectroscopic observations spanning 
multiple epochs (preferably with higher spatial resolution) are needed to confirm 
this interpretation.

Continued photometric and spectroscopic monitoring of this object can be used
to measure a time delay to compare with the value derived here, which
constrains the Hubble constant \citep{refsdal64}.
Multiple spectroscopic epochs can also be used 
to look for variation of the quasar emission line equivalent widths
consistent with a microlensing signature \citep[e.g.,][]{richards+04a}, which
constrains the size of the broad line region. 
Lensing can also distort the host galaxy image away from the bright active 
nucleus such that it can be resolved \citep{peng+06}. Probing the host galaxy 
of this object with HST imaging has the potential to allow the relationship 
between black hole mass and bulge mass to be examined for the most distant 
lensed quasar currently known.

This is the third high-redshift lensed quasar found in the spectrum of an 
object targeted by the SDSS as an LRG, joining the $z=3.6$ lensed quasar 
reported by \citet{johnston+03} and a $z=2.2$ lens from the SQLS 
high-redshift sample \citep{inada+09}. The SDSS-III\footnote{http://www.sdss3.org/} 
will target 1.5 million LRGs to a flux limit $\sim1.5$ mag fainter than the 
SDSS-I/II; this fifteen-fold increase  in the number of LRG spectra has the 
potential to result in a similar increase in discoveries of high-redshift quasars 
lensed by those galaxies and revealed by their spectra.

Another approach to identify quasars is by their variability 
\citep[e.g.,][]{schmidt+10,kozlowski+10}. Planned synoptic surveys 
such as PAN-STARRS \citep{panstarrs} and LSST \citep{lsst} will be able to 
identify high redshift quasar lenses through systematic searches for pairs of 
stellar objects at small separations that both vary (with a time delay), or 
by looking for resolved objects (galaxies) that vary \citep{kochanek+06}. 
\citet{om10} predict that thousands of new lensed quasars will be discovered 
by upcoming time-domain surveys, including several hundred at $z>4$ from the 
LSST alone.

\section{Acknowledgements}
We thank Jessica Evans and Chris Churchill for providing SPIcam flat fields.
We also thank the anonymous referee for informative comments which improved
the manuscript.
IDM, XF and FB acknowledge support from a Packard Fellowship for Science and 
Engineering and NSF grant AST 08-06861. 
PBH acknowledges support from NSERC.
N.~I. acknowledges support from MEXT KAKENHI 21740151.
MAS acknowledges support from NSF grant AST-0707266. 
DPS acknowledges support from NSF grant AST-0607634.

Use of the UH 2.2-m telescope for the observations is supported by NAOJ.

Funding for the SDSS and SDSS-II has been provided by the Alfred P. Sloan Foundation, the Participating Institutions, the National Science Foundation, the U.S. Department of Energy, the National Aeronautics and Space Administration, the Japanese Monbukagakusho, the Max Planck Society, and the Higher Education Funding Council for England. The SDSS Web Site is http://www.sdss.org/.

The SDSS is managed by the Astrophysical Research Consortium for the Participating Institutions. The Participating Institutions are the American Museum of Natural History, Astrophysical Institute Potsdam, University of Basel, University of Cambridge, Case Western Reserve University, University of Chicago, Drexel University, Fermilab, the Institute for Advanced Study, the Japan Participation Group, Johns Hopkins University, the Joint Institute for Nuclear Astrophysics, the Kavli Institute for Particle Astrophysics and Cosmology, the Korean Scientist Group, the Chinese Academy of Sciences (LAMOST), Los Alamos National Laboratory, the Max-Planck-Institute for Astronomy (MPIA), the Max-Planck-Institute for Astrophysics (MPA), New Mexico State University, Ohio State University, University of Pittsburgh, University of Portsmouth, Princeton University, the United States Naval Observatory, and the University of Washington.

{\it Facilities:} \facility{ARC (SPIcam), \facility{MMT (Red channel)}, \facility{UH:2.2m (Tek2k)}, \facility{Sloan} }

\end{document}